%
%
%
%
\font\bigg=cmbx10 at 17.3 truept
 at 12 truept


\font\twelverm=cmr10 scaled 1200    \font\twelvei=cmmi10 scaled 1200
\font\twelvesy=cmsy10 scaled 1200   \font\twelveex=cmex10 scaled 1200
\font\twelvebf=cmbx10 scaled 1200   \font\twelvesl=cmsl10 scaled 1200
\font\twelvett=cmtt10 scaled 1200   \font\twelveit=cmti10 scaled 1200

\skewchar\twelvei='177   \skewchar\twelvesy='60


\def\twelvepoint{\normalbaselineskip=12.4pt
  \abovedisplayskip 12.4pt plus 3pt minus 9pt
  \belowdisplayskip 12.4pt plus 3pt minus 9pt
  \abovedisplayshortskip 0pt plus 3pt
  \belowdisplayshortskip 7.2pt plus 3pt minus 4pt
  \smallskipamount=3.6pt plus1.2pt minus1.2pt
  \medskipamount=7.2pt plus2.4pt minus2.4pt
  \bigskipamount=14.4pt plus4.8pt minus4.8pt
  \def\rm{\fam0\twelverm}          \def\it{\fam\itfam\twelveit}%
  \def\sl{\fam\slfam\twelvesl}     \def\bf{\fam\bffam\twelvebf}%
  \def\mit{\fam 1}                 \def\cal{\fam 2}%
  \def\tt{\twelvett}
  \textfont0=\twelverm   \scriptfont0=\tenrm   \scriptscriptfont0=\sevenrm
  \textfont1=\twelvei    \scriptfont1=\teni    \scriptscriptfont1=\seveni
  \textfont2=\twelvesy   \scriptfont2=\tensy   \scriptscriptfont2=\sevensy
  \textfont3=\twelveex   \scriptfont3=\twelveex  \scriptscriptfont3=\twelveex
  \textfont\itfam=\twelveit
  \textfont\slfam=\twelvesl
  \textfont\bffam=\twelvebf \scriptfont\bffam=\tenbf
  \scriptscriptfont\bffam=\sevenbf
  \normalbaselines\rm}



\def\beginparmode{\endmode
  \begingroup \def\endmode{\par\endgroup}}
\let\endmode=\par
{\obeylines\gdef\
{}}
\def\singlespace{\baselineskip=\normalbaselineskip}

\def\oneandahalfspace{\baselineskip=\normalbaselineskip
  \multiply\baselineskip by 3 \divide\baselineskip by 2}

\newcount\firstpageno
\firstpageno=2
\footline={\ifnum\pageno<\firstpageno{\hfil}\else{\hfil\twelverm\folio\hfil}\fi}
\let\rawfootnote=\footnote              
\def\footnote#1#2{{\rm\singlespace\parindent=0pt\rawfootnote{#1}{#2}}}
\def\raggedcenter{\leftskip=4em plus 12em \rightskip=\leftskip
  \parindent=0pt \parfillskip=0pt \spaceskip=.3333em \xspaceskip=.5em
  \pretolerance=9999 \tolerance=9999
  \hyphenpenalty=9999 \exhyphenpenalty=9999 }
\def\dateline{\rightline{\ifcase\month\or
  January\or February\or March\or April\or May\or June\or
  July\or August\or September\or October\or November\or December\fi
  \space\number\year}}
\def\received{\vskip 3pt plus 0.2fill
 \centerline{\sl (Received\space\ifcase\month\or
  January\or February\or March\or April\or May\or June\or
  July\or August\or September\or October\or November\or December\fi
  \qquad, \number\year)}}


\hsize=6.5truein
\hoffset=.1truein
\vsize=8.5truein
\voffset=.05truein
\parskip=\medskipamount
\twelvepoint            
\oneandahalfspace       
\overfullrule=0pt       

\def\preprintno#1{
 \rightline{\rm #1}}    

\def\head#1{                    
  \filbreak\vskip 0.5truein     
  {\immediate\write16{#1}
   \raggedcenter \uppercase{#1}\par}
   \nobreak\vskip 0.25truein\nobreak}

\def\references                 
  {\head{References}            
   \beginparmode
   \frenchspacing \parindent=0pt \leftskip=1truecm
   \parskip=8pt plus 3pt \everypar{\hangindent=\parindent}}

\def\frac#1#2{{\textstyle{#1 \over #2}}}
\def\square{\kern1pt\vbox{\hrule height 1.2pt\hbox{\vrule width 1.2pt\hskip 3pt
   \vbox{\vskip 6pt}\hskip 3pt\vrule width 0.6pt}\hrule height 0.6pt}\kern1pt}
\newcount\pagenumber
\newcount\sectionnumber
\newcount\appendixnumber
\newcount\equationnumber

\newcount\citationnumber
\global\citationnumber=1

\def\ifundefined#1{\expandafter\ifx\csname#1\endcsname\relax}
\def\cite#1{\ifundefined{#1} {\bf ?.?}\message{#1 not yet defined,}
\else \csname#1\endcsname \fi}

\def\docref#1{\ifundefined{#1} {\bf ?.?}\message{#1 not yet defined,}
\else \csname#1\endcsname \fi}

\def\article{
\def\eqlabel##1{\edef##1{\sectionlabel.\the\equationnumber}}
\def\seclabel##1{\edef##1{\sectionlabel}}
\def\citelabel##1{\edef##1{\the\citationnumber}{\global\advance\citationnumber
by1}}
}

\def\appendixlabel{\ifcase\appendixnumber\or A\or B\or C\or D\or E\or
F\or G\or H\or I\or J\or K\or L\or M\or N\or O\or P\or Q\or R\or S\or
T\or U\or V\or W\or X\or Y\or Z\fi}

\def\sectionlabel{\ifnum\appendixnumber>0 \appendixlabel
\else\the\sectionnumber\fi}

\def\beginsection #1
 {{\global\subsecnumber=1\global\appendixnumber=0\global\advance\sectionnumber
by1}\equationnumber=1
\par\vskip 0.8\baselineskip plus 0.8\baselineskip
 minus 0.8\baselineskip
\noindent $\S$ {\bf \sectionlabel. #1}
\par\penalty 10000\vskip 0.6\baselineskip plus 0.8\baselineskip
minus 0.6\baselineskip \noindent}

\newcount\subsecnumber
\global\subsecnumber=1

\def\subsec #1 {\bf\par\vskip8truept  minus 8truept
\noindent \ifnum\appendixnumber=0 $\S\S\;$\else\fi
$\bf\sectionlabel.\the\subsecnumber$ #1
\global\advance\subsecnumber by1
\rm\par\penalty 10000\vskip6truept  minus 6truept\noindent}

\def\beginappendix #1
{{\global\subsecnumber=1\global\advance\appendixnumber
by1}\equationnumber=1\par
\vskip 0.8\baselineskip plus 0.8\baselineskip
 minus 0.8\baselineskip
\noindent
{\bf Appendix \appendixlabel . #1}
\par\vskip 0.8\baselineskip plus 0.8\baselineskip
 minus 0.8\baselineskip
\noindent}

\def\no{\eqno({\rm\sectionlabel}
.\the\equationnumber){\global\advance\equationnumber by1}}

\def\ref #1{{\bf [#1]}}

\article

\def\cS{{\cal S}}
\def\vf{\varphi}
\def\D{{\cal D}}
\def\P{{\cal P}}
\def\M{{\cal M}}
\def\l{\lambda}
\def\vf{\varphi}

\vskip -0.8truein
\preprintno{DIAS-STP-95-36, ITFA-94-36}
\preprintno{hep-th/9511090}
\vskip 20 true pt
\centerline{\bf\bigg Field Theory Entropy, the $H$-theorem and}
\centerline{\bf\bigg the Renormalization Group} 
\vskip 20 true pt
\centerline{Jos{\'e} Gaite}
\centerline{\it Instituto de Matem{\'a}ticas y F{\'\i}sica Fundamental}
\centerline{\it C.S.I.C., Serrano 123,}
\centerline{\it 28006 Madrid, Spain}
\centerline{and}
\centerline{Denjoe O'Connor}
\centerline{\it School of Theoretical Physics,}
\centerline{\it D.I.A.S., 10 Burlington Rd.,}
\centerline{\it Dublin 4, Ireland}
\vskip 20 true pt
\centerline{\bf Abstract}
We consider entropy and relative entropy in Field theory and establish
relevant monotonicity properties with respect to the couplings.
The relative entropy in a field theory with a
hierarchy of renormalization group fixed points ranks the fixed points,
the lowest relative entropy being assigned to the highest multicritical point.
We argue that as a consequence of a generalized $H$ theorem
Wilsonian RG flows induce an increase in entropy and propose
the relative entropy as the natural quantity which increases
from one fixed point to another in more than two dimensions.
\vskip 20 true pt
PACS codes: 11.10.-z, 65.50.+m, 5.70.Jk, 64.60.Ak
\vfill\eject

\beginsection{Introduction}
The concept of entropy was introduced by Clausius through the study of
thermodynamical systems. However it was Boltzmann's essential discovery that
entropy is the natural quantity that bridges the microscopic and macroscopic
descriptions of a system which gave it its modern interpretation.
A more general definition, proposed by Gibbs
allowed its extension to any system where probability theory
plays a r{\^o}le. It is a variant of this entropy which we discuss in a
field theoretic context. Boltzmann also defined, in kinetic theory, a
quantity $H$, that decreases with time and for a non-interacting gas
coincides with the entropy at equilibrium ($H$-theorem). These ideas also
admit generalization and in our context we
will see that analogous ``non-equilibrium'' ideas can
be associated with Wilsonian renormalization in our field theory
entropic setting.

Probabilistic entropy can be defined for a field theory and in terms
of appropriate variables is either a monotonic
or convex function of those variables. A variant of it, the
relative entropy is suited to the study of systems where there is a
distinguished point as in the case of critical phenomena, where a critical
point is distinguished.

We shall see that monotonicity of the relative entropy along lines that
depart from the distinguished point in coupling space entails its increase
in the crossover from the critical behavior associated with one domain of
scale invariance or fixed point to that associated with a ``lower''
fixed point, thus providing a quantity that naturally ``ranks'' the
fixed points. This property is a consequence of convexity of
the appropriate thermodynamic surface, which in turn is reflected in the
general structure of the phase diagram \citelabel{\Fisher}[\cite{Fisher}].
The phase diagrams of lower critical points emerge as projections
of the larger phase diagram. We shall see that the natural geometrical
setting for these phase diagrams is projective geometry.

There have been many attempts to capture the irreversible nature of a
Wilson renormalization group (RG) flow in some function which
is intended to be monotonic under the iteration of a Wilson
RG transformation \citelabel{\various}[\cite{various}].
These attempts have been successful in two dimensions where
the Zamolodchikov $C$ function has the desired property.
The monotonicity of the flow of the $C$-function under scale
transformations is reminiscent of Boltzmann's $H$-function and this
result has been accordingly called the $C$ theorem. Boltzmann's $H$
function was the generalization of entropy to non-equilibrium situations,
in particular, to a gas with an arbitrary particle distribution in phase
space. He proved that $H$ increases whenever the gas evolves to its
Maxwell-Boltzmann equilibrium distribution \citelabel{\Balian}
[\cite{Balian}], effectively
making this evolution an irreversible process. We will argue that
an analogue ``non-equilibrium'' probabilistic entropy  for a field theory
provides a natural function that must increase under a Wilsonian RG flow.
We shall consider a version of the $H$ theorem suited to our needs, to see
how the increase occurs. A differential increase along the RG trajectories
demands detailed knowledge of the flow lines; however,
statements about the ends of the flows are more robust and thus more
easily established. It is such statements that we shall establish.

Among other attempts to apply the methods of
entropy and irreversibility to Quantum Field Theory, it was
shown in \citelabel{\Merca}[\cite{Merca}] that an entropy defined from the
quantum particle density, understood as a probability density, should
increase as the field theory reaches its classical limit. If we regard
this limit as a crossover between different theories, that result should
be directly connected to ours. Regarding the connection with two dimensional
conformal field theories and Zamolodchikov's $C$ theorem
it is noteworthy that calculations of the geometrical or
entanglement entropy (see \citelabel{\BombSorkSred}[\cite{BombSorkSred}]
for background) give a quantity proportional to the central charge $c$
\citelabel{\Hol}[\cite{Hol}],
we will not however pursue possible connections with the entanglement
entropy here.

The structure of the paper is as follows: In section 2 we review the
definitions of entropy and relative entropy and adapt them to field theory.
We study some of their properties, especially the property of monotonicity
with respect to couplings, related with convexity. Section 3
discusses the crossover of the relative entropy between field theories.
We provide some examples, ranging from the trivial crossover, in the
Gaussian model as a function of mass, to the tricritical to critical
crossover, which illustrates the generic features of this phenomenon.
This section ends with
a brief study of the geometric structure of phase diagrams relevant to
crossover phenomena. Although section 3 heavily relies on RG constructs,
the picture of the RG used is somewhat simple minded. In section 4 we
improve on that picture, introducing Wilson's RG ideas. We see how these
ideas naturally lead one to interpret crossover from {\it cutoff
dependent} to {\it cutoff independent} degrees of freedom as an
irreversible process in the
sense of Thermodynamics and therefore to consider a non-equilibrium
field-theoretic $H$ theorem type entropy.

\beginsection{Entropy in Field Theory, definition and properties}
For a normalized probability distribution $\P$,
we take as
our definition of probabilistic entropy,
\eqlabel{\Entropy}
$$\cS_a=-{\rm Tr}\,\P\ln\P\no$$
and will refer to this as ``absolute probabilistic entropy''.
For example for a single random variable $\phi$
governed by the normalized Gaussian probability distribution
\eqlabel{\Gauss}
$$\P=e^{-{1\over2}m^2\phi^2-j\phi+W[j,m^2]}\no$$
where $W[j,m^2]=-{j^2\over2m^2}+{1\over2}\ln {m^2\over2\pi}$
and ${\rm Tr}$ is understood to mean integration over $\phi$.
The absolute probabilistic entropy is given by
\eqlabel{\absgauss}
$$\cS_a={1\over2}-{1\over2}\ln{m^2\over2\pi}\no$$
A natural generalization of this entropy known as the relative entropy
\citelabel{\Ellis}[\cite{Ellis}] is given by
\eqlabel{\RelEntropy}
$$\cS[\P,\P_0]={\rm Tr}[\P\ln(\P/\P_0)]\no$$
where $\P_0$ specifies the a priori probabilities. The sign change
relative to (\docref{Entropy}) is conventional. Relative entropy plays an
important r{\^o}le in statistics and the theory of
large deviations\citelabel{\LewisPfisterSullivan}\citelabel{\LewisPfister}
[\cite{LewisPfisterSullivan},\cite{LewisPfister}].
It is a convex function of $\P$ with $\cS\ge0$ and equality applying iff
$\P=\P_0$.
It measures the statistical distance between the probability distributions
$\P$ and $\P_0$ in the sense that the smaller $\cS[\P,\P_0]$ the
harder it is to discriminate between
$\P$ and $\P_0$. The infinitesimal form of this distance provides a
metric known as the Fisher information matrix
\citelabel{\Amari}[\cite{Amari}] and provides
a curved metric on the space of parameterized probability
distributions and the space of couplings in field theory
\citelabel{\OConSte}[\cite{OConSte}].
For example if we consider the probability distribution (\docref{Gauss}),
with $j=0$, for simplicity, the entropy of the Gaussian distribution with
standard deviation $m^2$ relative to the Gaussian distribution with
standard deviation $m_0^2$ is given by
\eqlabel{\relgauss}
$$\cS[m^2,m_0^2]={1\over2}\ln{m^2\over m_0^2}+{m_0^2\over2m^2}-{1\over2}\no$$
and can be easily seen to have the desired properties.
By taking the a priori probabilities to be given by the
uniform distribution we recover (\docref{Entropy}), modulo a sign.
However, we see that (\docref{relgauss})  approaches
(\docref{absgauss}) but modulo a divergent constant as
$m_0\rightarrow0$. This reflects the fact that the uniform
distribution is not normalizable. The uniform distribution in this
setting doesn't strictly fit the criteria of a suitable a priori
distribution $\P_0$ and therefore violates the
assumptions guaranteeing the positivity of the relative entropy.
More generally for a continuously distributed random variable a more suitable
distribution, with respect to which one can define the a priori
probabilities, is one that resides in the same function space.

In the case of a field theory ${\rm Tr}$ will be a path integral over
the field configurations and just as when defining the partition
function of a field theory an ultraviolet and an infrared regulator are,
in general, necessary.
Convenient infrared regulators will be to consider a massive field
theory in a finite box. It is then convenient to deal with the entropy
per unit volume or specific
entropy $S=\cS/V$ where $V$ is the volume of the manifold, $\M$, on which the
field theory is defined. One would generally expect that $S$ would contain
divergent contributions as the regulators are removed. However,
these contributions disappear in an appropriately defined relative entropy.

For a field theory consider
\eqlabel{\pofz}
$$\P_z={\rm e}^{-I^0[\phi,\{\l\}]-z\,I^c[\phi,\{l\}]+W[z,\{\l\},\{l\}]}\no$$
where $W[z,\{\l\},\{l\}]=-\ln Z[z,\{\l\},\{l\}]$, with
$$Z[z,\{\l\},\{l\}]=\int\D[\phi]\,{\rm
e}^{-I^0[\phi,\{\l\}]-zI^c[\phi,\{l\}]}\no$$
i.e. the total action  for the random field variable
$\phi$ is given by $I=I^0[\phi,\{\l\}]+z I^c[\phi,\{l\}]$.
We have divided the parameters of the theory into two sets:
The set $\{\l\}$ is the set of  coupling constants associated with the
fixed distribution $\P_0$ and $\{l\}$ are those associated with the
additional, or crossover, contribution to the action $z I^c$.
The two sets are assumed to be distinct, the set $\{l\}$ may, however,
incorporate  changes to the couplings of the
set $\{\l\}$.

We have introduced the variable $z$ primarily for later convenience.
For a given functional integral ``measure'', associated with integration over a
fixed
function space (this may be made well defined by
fixing for example ultraviolet and infrared cutoffs),
$W[z,\{\l\},\{l\}]$ reduces to $W^0[\{\l\}]$ when $z=0$.
With the notation
$$\langle X \rangle=\int\D[\phi]\,X[\phi]\,{\rm e}^{-I^0[\phi,\{\l\}]-z
I^c[\phi,\{l\}]+W[z,\{\l\},\{l\}]} \no$$
assuming analyticity in $z$ in the neighborhood of $z=1$,
the value of principal interest to us, we have
$${d W[z,\{\l\},\{l\}]\over dz}=\langle I^c\rangle\no$$
and more generally
$${d\langle X\rangle\over dz} = -\left(\langle X I^c\rangle -\langle
X\rangle\langle I^c\rangle\right)$$
We can therefore express the relative entropy as
$$\cS[z,\{\l\},\{l\}]=
W[z,\{\l\},\{l\}]-W^0[\{\l\}]-z\langle I^c[\phi,\{l\}]\rangle\no$$
It is, the Legendre transform with
respect to $z$ of $W^c=W-W^0$:
\eqlabel{\Legtransform}
$$\cS[z,\{\l\},\{l\}]=W^c[z,\{\l\},\{l\}]-z{dW^c[z,\{\l\},\{l\}]\over dz}\no.$$
Next consider the derivative with respect to $z$ of $\cS$.
\eqlabel{\wderiv}
$${d\cS[z,\{\l\},\{l\}]\over dz}=-z{d^2W[z,\{\l\},\{l\}]\over dz^2}\no$$
Re-expressing this in terms of expectation values we have
\eqlabel{\monotonicent}
$$z{d\cS[z,\{\l\},\{l\}]\over dz}=z^2\langle (I^c -\langle
I^c\rangle)^2\rangle\no$$
implying that $\cS$ is a monotonic increasing function of $|z|$ which
is zero at $z=0$. We also deduce from (\docref{wderiv}) and
(\docref{monotonicent})
that $W$ is a convex function of $z$.

Note that the expression (\docref{Legtransform}) is amenable to standard
treatment
by field theoretic means. In perturbation theory, it is diagrammatically
a sum of connected vacuum graphs.
Furthermore, if the action is a linear combination of terms
\eqlabel{\linIc}
$$I^c[\phi,\{l\}]=l^af_a[\phi]\no$$
then with $zl^a=t^a$ ($z$ is an overall factor)
we have
\eqlabel{\entropyleg}
$$\cS[\{\l\},\{t\}]=W[\{\l\},\{t\}]-W[0]-t^a\partial_a W[\{\l\},\{t\}]\no$$
where $\partial_a={\partial \over\partial t^a}$.
Thus for this situation the relative entropy of the field theory is the
complete Legendre
transform of the generating function $W$ with respect to all the
couplings $t^a$. The negative of the ``absolute" entropy or
entropy relative to the uniform distribution
(equivalent to $I^0[\phi,\{\l\}]=0$) would be the complete Legendre
transform with respect to all the couplings
in such a field theory.  In terms of its natural variables
 $\langle f_a \rangle =
{\partial_a W}$ the relative entropy itself is a convex function (see below).
It proves useful in what follows to regard it as a function of
the couplings through $\langle f_a \rangle(t)$.

Let us consider the change in relative entropy due to an infinitesimal
change in the couplings of the theory. This can be expressed as
a 1-form on the space of couplings.  A little re-arrangement shows that
such a change can be expressed in the form
\eqlabel{\intfactor}
$$d\cS=z(d\langle I^c\rangle -\langle dI^c\rangle)\no$$
which implies that $z^{-1}$ performs the r{\^o}le of an integrating factor
for the difference of infinitesimals $d\langle I^c\rangle-\langle dI^c\rangle$,
just as temperature does for the absolute entropy.
We could more generally consider different
$z$'s for each of the composite operators $f_a[\phi]$
and obtain the generalization of (\docref{intfactor})
$$d\cS=\sum_{a}Z_{f_{a}}
(d\langle f_a[\phi]\rangle-\langle df_a[\phi]\rangle)$$
In renormalization theory the $Z_{f_{a}}$ play the r{\^o}le of composite
operator renormalizations (e.g. $l^af_a[\phi]={1\over2}\int t\,\phi^2$
the composite operator $\phi^2$ gets renormalized by $Z_{\phi^2}$).
Thus one could interpret composite operator renormalization factors $Z_{f_{a}}$
(or in the example $Z_{\phi^2}$) as integrating factors.

Again for the case (\docref{linIc})  since
$$z^2 \langle (I^c-\langle I^c\rangle)^2\rangle =
t^a\langle (f_a-\langle f_a\rangle)(f_b-\langle f_b\rangle)\rangle t^b\no$$
and each of the $l^a$ are arbitrary, we see that the quadratic form
$$Q_{ab}=\langle (f_a-\langle f_a\rangle)(f_b-\langle f_b\rangle)\rangle =
-{\partial^2 W \over{\partial t^a \partial t^b}}\no$$ is a
positive definite matrix.
This establishes the key property that $W$ is a convex function of the
couplings. $\cS$ is similarly a convex function of the
$\langle f_a\rangle$,  since
$$Q^{ab}= Q_{ab}^{-1} = {\partial^2 \cS \over
{\partial \langle f_a\rangle \partial \langle f_b\rangle}}.\no$$
The matrix $Q_{ab}$ is the Fisher information matrix and plays the r{\^o}le of
a
natural metric on the space of couplings $\{l\}$ measuring the infinitesimal
distance between probability distributions.

We end this section by emphasizing that in the above we have established that
$W$ is a convex function of the $l^a$ and $\cS$ is a convex function of the
$\langle f_a\rangle$.  Note that: {\it the usual effective action can be
viewed as the relative entropy} with $z I^c[\phi,\{l\}]=\int_{\M} J\phi$
and is therefore a convex function
of $\langle\phi\rangle$. The relative entropy is equivalently a generalization
of the effective action to a more general setting.
A final observation is that the relations
\eqlabel{\equilconstr}
$$\bar f_a=\langle f_a\rangle={\partial_a W}(t)\no$$
are our field equations (on-shell conditions) and can be associated with
equilibrium.
If one releases these constraints by for example leaving
the equilibrium setting,
one can consider $\cS$ as a function of both the $\bar f_a$ and $l^a$.
The equilibrium conditions are then specified by (\docref{equilconstr}).

\beginsection{Crossover between Field Theories}
The concept of crossover arises in the physics of phase transitions,
where it means the change from one
type of critical behavior to another. This implies a change of critical
exponents or any other quantity associated with critical behavior.
In our context, a field theory (FT) is defined by a Lagrangian with
a number of coupling constants. We will restrict our considerations to the
case of super-renormalizable theories, in which case the theories can be taken
to provide well defined microscopic theories.
The Lagrangian captures the universality class of a particular phase transition
when the relevant couplings are tuned to appropriate values; these relevant
couplings
constitute a parameterization of the space of fields and couplings close to the
associated fixed point (FP) of the RG.
The functional integral provides global information, which can be depicted
in a phase diagram, with variables $W, \{l\}$.
The most unstable FP will therefore have the largest dimensional phase
diagram and far from this FP may exist another where one (or more) of
the maximal set of couplings becomes
irrelevant\footnote{*}{Here relevant and irrelevant have both their intuitive
and RG meaning.}
and drops out.
This implies the change to a universality class with fewer relevant couplings,
hence a reduced phase diagram corresponding to projecting out the
couplings which became irrelevant.
The second FP and the reduced phase diagram define a new field theory.

It is fairly easy to see that in the region where homogeneous scaling holds and
the
RG trajectories satisfy linear RG equations there can be no more
fixed points.  One can define new coordinates called
non-linear scaling fields \citelabel{\Wegner}[\cite{Wegner}] where homogeneous
scaling applies throughout the phase diagram.
This possibility is also well known in the theory of
ordinary differential equations (ODEs), where it is called Poincar{\'e}'s
theorem \citelabel{\Arnold}[\cite{Arnold}, pg.\ 175].
In these co-ordinates, then, any other FP must be placed at infinity in
a coordinate system adapted to the first FP.
To study the crossover, when a FP is at infinity, we need to perform
some kind of compactification of the phase diagram.
Thus, we shall think of the total phase diagram as a compact manifold
containing the maximum number of {\it generic} RG FP.
This point of view is especially sensible regarding the topological
nature of RG flows. Furthermore, thinking
of the RG as just an  ODE indicates
what type of compactification of phase diagrams is adequate:
It is know in the theory of ODEs that the analysis of the
flow at infinity and its possible singularities can be done
by completing the affine space to
projective space \citelabel{\Lefschetz}[\cite{Lefschetz}]. This as we shall
see is also appropriate for phase diagrams.

We will restrict our considerations in what follows to
scalar $Z_2$ symmetric field theories with polynomial potentials and
non symmetry breaking fields.  For illustration, we
will discuss some exact results pertaining to solvable statistical models,
which illuminate the behaviour of  the field theories in the same universality
classes.

\subsec{Case (0), The Gaussian model and the zero to infinite mass crossover}
Consider the action
$$I_0^0[\phi,\{\l(0)\}]
=\int_{\M}\left\{{\alpha\over2}(\partial\phi)^2+{r_c\over2}\phi^2\right\}\no$$
The action associated with $\P_z$ is then
$$I_0[\phi,\{\l(0)\},t]=I_0^0[\phi,\{\l(0)\}]+\int_{\M}{t\over2}\phi^2\no$$
The crossover here is that associated with $z=t$. The model is pathological
in that it is not well defined for $t<0$ where there is no ground state,
but our interest is in $t\ge0$. The crossover of interest here is then
from $t=0$ to large values of $t$.
To make the model completely well defined we
place it on a lattice and take the continuum limit.

For the Gaussian model on a square lattice with lattice spacing,
taken for simplicity to be $a\sqrt{\alpha}$, and
with periodic boundary conditions and sides of length
$L=Ka\sqrt{\alpha}$, in $d$ dimensions,  we have
in the thermodynamic limit $K\rightarrow\infty$%
\citelabel{\BerlinKac}[\cite{BerlinKac}] 
$$W[a,r]={K^d\over
2}\int_{-\pi}^{\pi}{d\omega_1\over2\pi}\cdots\int_{-\pi}^{\pi}{d\omega_d
\over 2\pi}\,\,
\ln\left\{{{4\over a^2}\sin^2({\omega_1\over 2})+\cdots
+{4\over a^2}\sin^2({\omega_d\over 2})+r_c+t\over 2\pi}\right\}\no$$
With the critical point of the model at $t=0$ we have $r_c=0$.
The relative entropy is
$$\cS[a,t]=W[a,t]-W[a,0]-t{dW[a,t]\over dt}\no$$
so if $W[a,t]$ took the form $W[a,t]=\tilde W[a,t]+c + b t$
the linear term $c+bt$ would not contribute to the relative entropy.
In the thermodynamic limit, if we restrict our considerations to a
dimensionally regularized continuum
model then for $d<4$ the divergences that require subtraction are indeed
of the linear form and we find
that the relative entropy per unit volume is given by
\eqlabel{\gaussrelent}
$$S=
{(d-2)\pi\over 2\sin({\pi(d+2)\over2})\Gamma({d+2\over2}){(4\pi)}^{d\over2}}
\,t^{d\over2}\no$$

For $d>2$ and sufficiently small $t$, in the neighbourhood of the critical
point, the relative entropy of both the continuum model and the lattice
model agree.
This can be seen by noting that the second derivative of $W$ with respect to
$t$ diverges for small $t$ and, for $d<4$, the coefficient of divergence is the
same
for both the lattice and continuum expressions. Thus integrating back to obtain
$W[t]$ will give expressions which differ by only a linear term in $t$ for
small $t$ but
this does not affect the relative entropy.
{}From  (\docref{gaussrelent}) the increase in relative
entropy with $t$ is manifest.

\subsec{Case (i), the Ising universality class}
Let us next consider the two dimensional Ising model on a rectangular lattice.
For simplicity we will restrict our considerations to equal couplings in the
different directions.  Since the random variables here (the Ising spins) take
discrete values it is natural to consider the absolute entropy
which corresponds to choosing entropy relative to the discrete counting
measure and a sign change.  This is the standard absolute entropy in this
case.  This model, as is well known, admits an exact solution
\citelabel{\Onsager}[\cite{Onsager}] for the partition function with
$$\eqalign{W[k] =& -(1/2) \ln(2 \sinh(2k)) -
\int_0^\pi{d\omega\over 2\pi}\cosh^{-1}\left[\cosh(2k)
\cosh(2K(k)) -
\cos(\omega) \right]}
\no$$
for a rectangular lattice where $K(k)= {1\over2}\ln\coth(k)$ and $k={J\over
k_BT}$.
The entropy is then
\eqlabel{\Onsagerentropy}
$$\cS_a=-\left(W(k)-k{dW(k)\over dk}\right)\no$$
and plotted against $k$ in figure 1a.
The monotonicity  of the entropy becomes one of
convexity when the entropy is expressed in terms of the internal
energy $U$ as can be seen in figure 1b.
\vfill\eject
\par\vskip\baselineskip
\input epsf
\epsfxsize=0.45\hsize
\centerline{\bf Fig. 1a: The Entropy $S_a(k)$ for the 2d Ising
Model} \par\vskip-2\baselineskip
\centerline{\epsffile{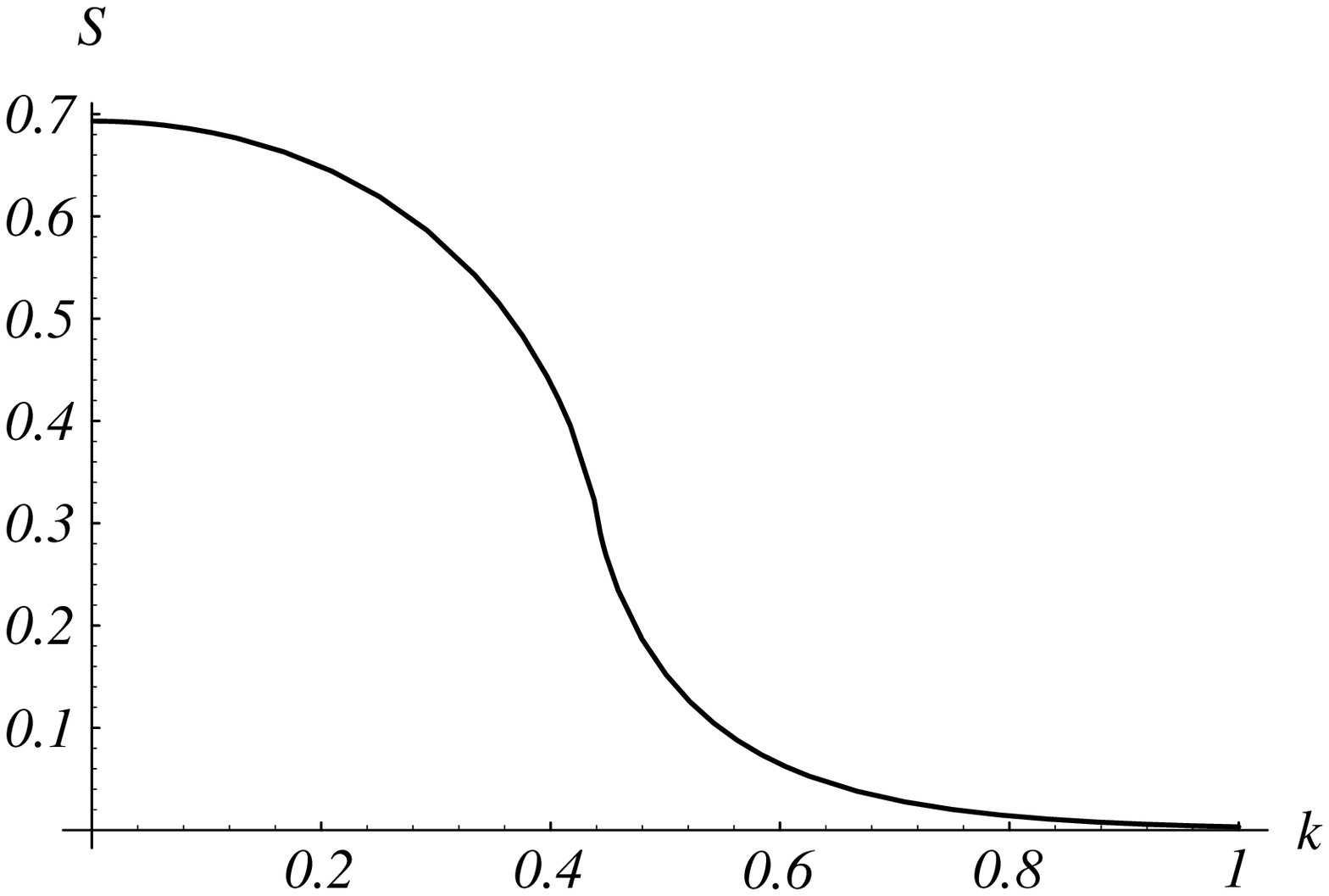}}
\par\vskip-2\baselineskip
\par\vskip\baselineskip
\input epsf
\epsfxsize=0.45\hsize
\centerline{\bf Fig. 1b: The Entropy $S_a(U)$ for the 2d Ising Model}
\par\vskip-2\baselineskip
\centerline{\epsffile{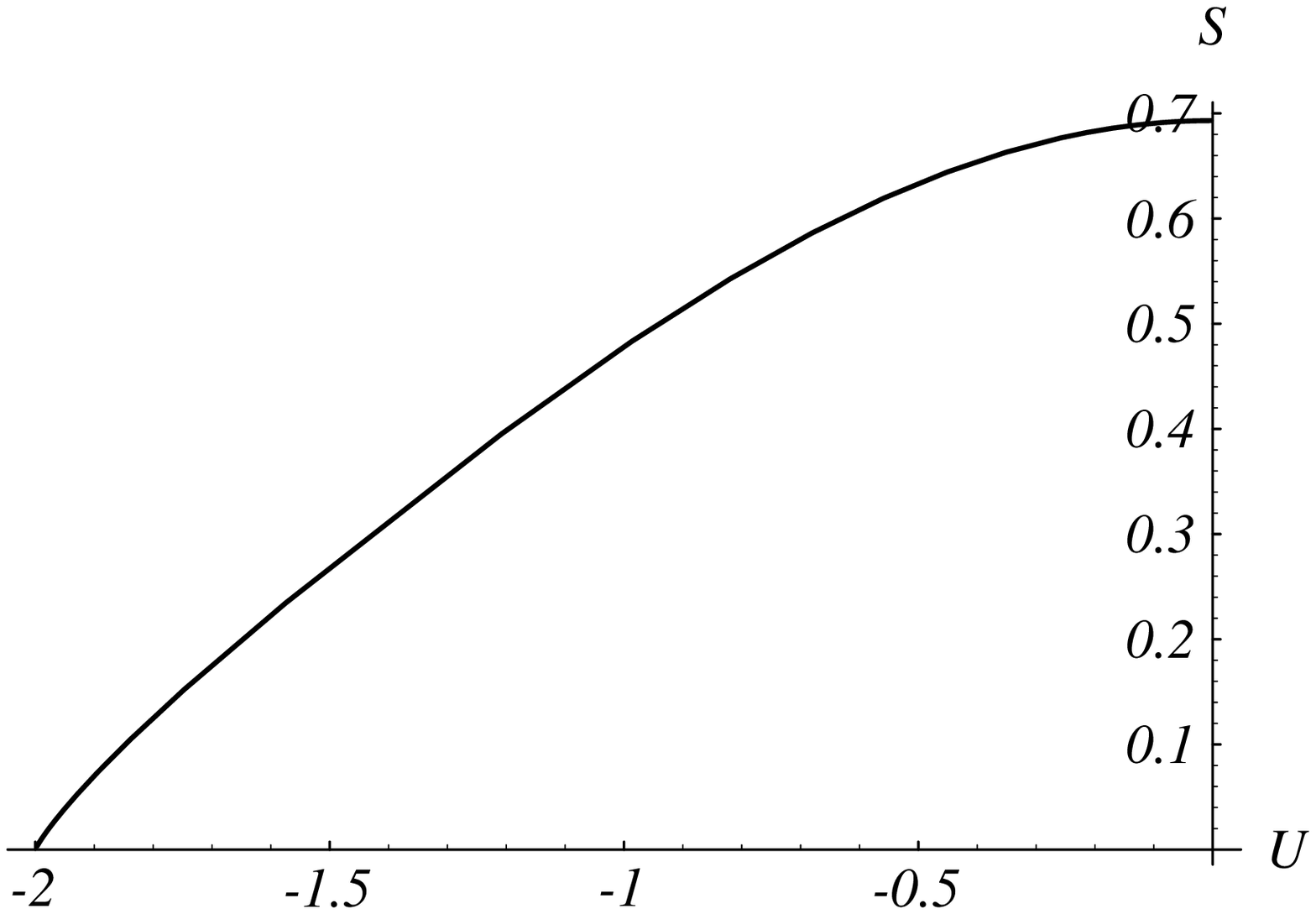}}
\par\vskip-2.5\baselineskip

Now, of course, we can also consider relative entropy in this setting.
Since near its critical point the two dimensional Ising
model is in the universality class of a $\phi^4$ field theory, to
facilitate
comparison with the field theory it is natural to choose entropy
relative to the critical point lattice Ising model. This is also natural
since the
critical point is a preferred point in the model. This relative entropy is
given by $$\cS=W(k)-W(k^*)-(k-k^*){dW(k)\over dk}\no$$
where $k^*={1\over2}\ln(\sqrt{2}+1)\sim 0.4406868$ is the
critical coupling of the Ising model.
We have plotted this in figure 2a. We see that it is a monotonic
increasing function of $|k-k^*|$ and is zero at the
critical point. In figure 2b we plot this entropy as a function
of the relevant expectation value, the internal energy
$U={dW\over dk}$,  and set the origin at  $U^*$,
the internal energy at the critical point. Naturally, the graph
is convex.
\par\vskip\baselineskip
\input epsf
\epsfxsize=0.45\hsize
\centerline{\bf Fig. 2a: The Relative Entropy $S(k,k^*)$ for the 2d Ising
Model} \par\vskip-2.5\baselineskip
\centerline{\epsffile{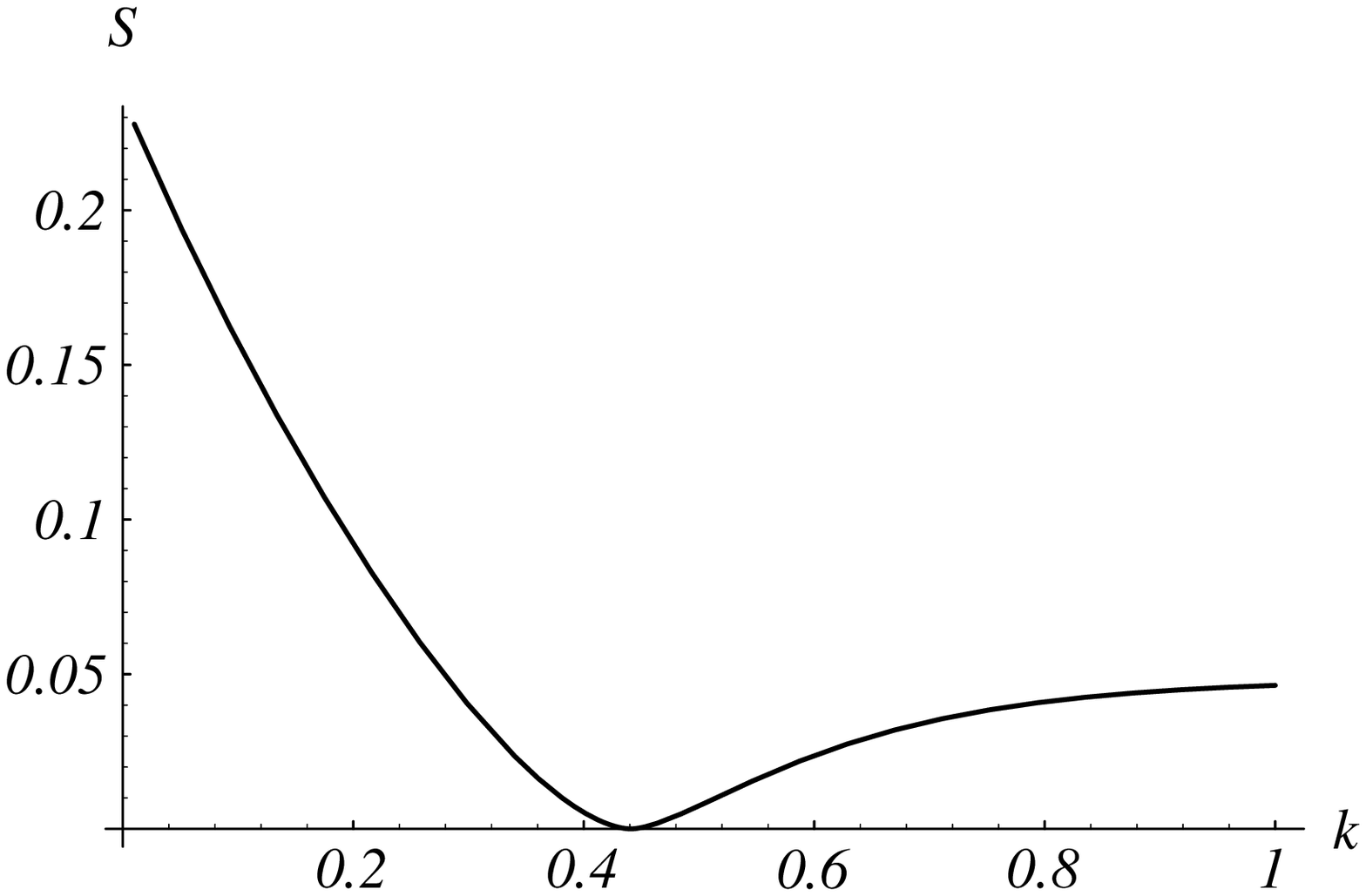}}
\par\vskip-2\baselineskip
\input epsf
\epsfxsize=0.45\hsize
\centerline{\bf Fig. 2b: The Relative Entropy $S(U,U^*)$ for the 2d Ising
Model}
\par\vskip-2\baselineskip
\centerline{\epsffile{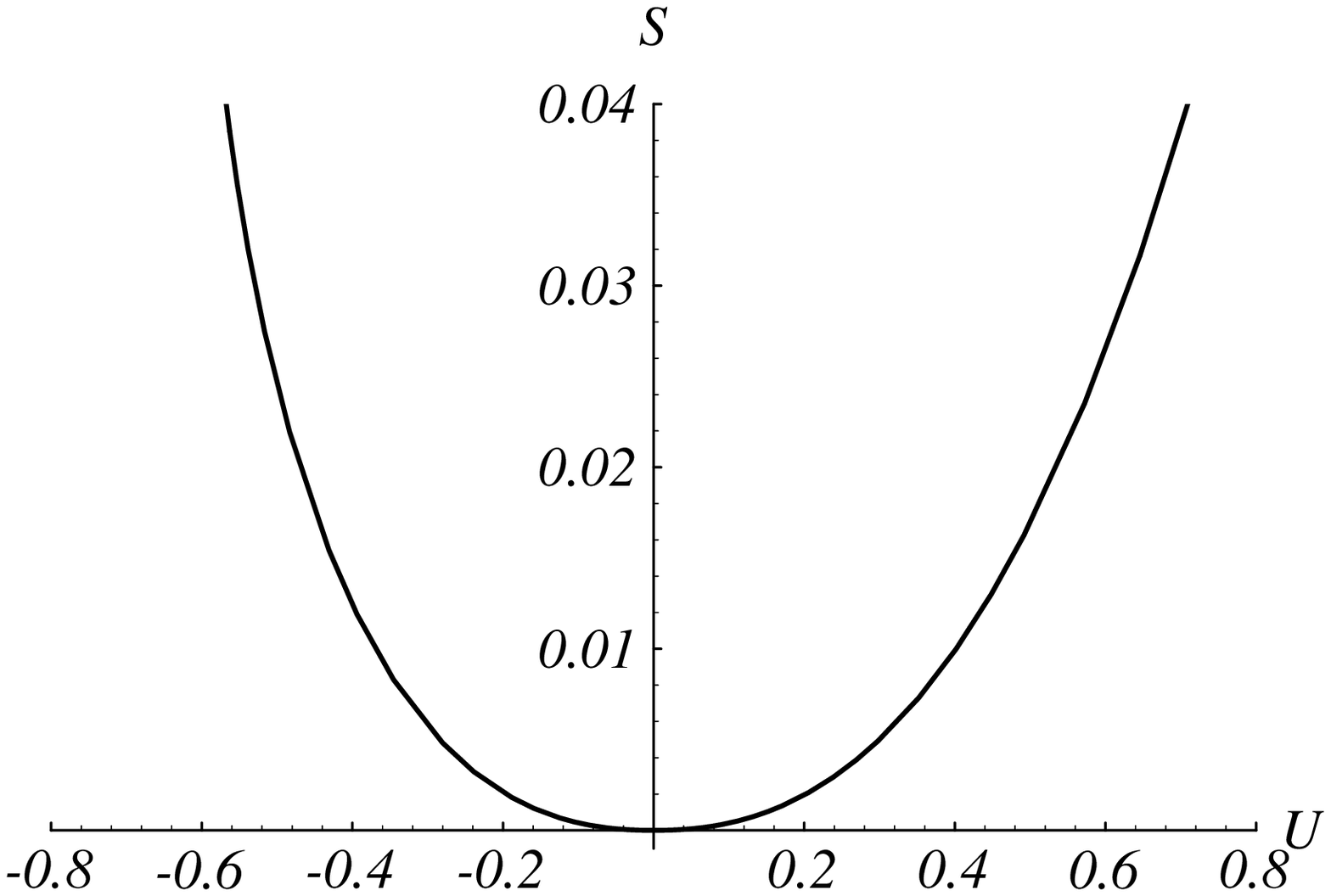}}
\par\vskip-2\baselineskip

In more than two dimensions the Ising model has not been solved exactly.
Its critical behaviour is in the universality class of a $\phi^4$ field
theory, so
we expect the general features of the two models to merge near the critical
point. We will next consider the $\phi^4$ theory.

We will choose the fixed probability distribution $\P_0$ for the $\phi^4$
theory
to be that associated with the critical point, or massless theory,  which
is described by the action
\eqlabel{\isingrelentropy}
$$I_1^0[\phi,\{\l(1)\}]
=\int_{\M}\left\{{\alpha\over2}(\partial\phi)^2+{r_c\over2}\phi^2+{\l\over
4!}\phi^4\right\}\no$$
with $\l$ some arbitrary but fixed value of the bare coupling constant. We
restrict our
considerations to $d<4$ where the theory is super-renormalizable. The parameter
$r_c$ depends on the cut-off (UV regulator) needed to render the theory at a
path
integral level well defined, and is chosen such that the correlation length is
infinite.
The complete action associated with $\P_z$ is
$$I_1[\phi,\{\l(1)\},t]=I_1^0[\phi,\{\l(1)\}]+\int_{\M}{t\over2}\,\phi^2\no$$
The crossover of interest here is that associated with
$z=|t|$. There are clearly two branches to
the crossover, that for $t$ positive, and negative respectively.
We will restrict our considerations to the positive branch,
corresponding
to  $\langle\phi\rangle=0$, and the range of $t$ is from $0$ to $\infty$.
The identification of $z$ with $t$ allows us to use the arguments of
the previous section. From (\docref{monotonicent}) we
conclude that the relative entropy is a
monotonic function along this crossover line. This is the crossover line from
the
Wilson Fisher fixed point to the infinite mass Gaussian fixed point.

In the presence of a fixed UV cutoff one could consider the
reference probability distribution
to be that for which $\l=0$ and then place $\l$ into the crossover portion of
the
action.  This provides us with another crossover and in this more complicated
phase diagram there are in fact two Gaussian fixed points;
a massless and infinite mass one,
both associated with $\l=0$ (see \citelabel{\NCS}[\cite{NCS}] for a
description of the total phase diagram).
The crossover between them is that associated with
``case (0)'' described above. If one further restricts to
$\l=\infty$, this is equivalent to restricting to the fixed point
coupling and is believed to be
equivalent to the Ising model in the scaling region. The parameters
$t$  and $k$ then should play equivalent r{\^o}les, and describe the
same crossover. In the $\phi^4$ model one can further consider
crossovers associated with varying $\l$ at fixed $t$, by including a term
$\int_{\M}{l\over 4!}\,\phi^4$ in $I^c$. In this family
there will be a crossover curve at infinity which varies from one infinite
mass Gaussian Fixed point to another. Such crossovers
can be viewed as a special case of the next example.

\subsec{Case (ii) Models with two crossover parameters}
Here the action for the fixed distribution from which we calculate
the relative entropy is taken to be
\eqlabel{\tricrelentropy}
$$I_2^0[\phi,\{\l(2)\}]
=\int_{\M}\left\{{\alpha\over2}(\partial\phi)^2
+{r_{tc}\over2}\phi^2+{\l_{tc}\over4!}\phi^4+{g\over 6!}\phi^6\right\}\no$$
($g$ fixed) and the action of the model is
$$I[\phi,\{\l(2)\},t,l]=I_2^0[\phi,\{\l(2)\}]
+\int_{\M}\left\{{t\over2}\phi^2+{l\over4!}\phi^4\right\}\no$$
The tricritical point corresponds to both $t$ and $l$  zero. There is now a
plane
to be considered. First consider the line formed setting $l=0$ and ranging
$t$ from zero to infinity. This is a line leaving the tricritical point and
going to
an infinite mass Gaussian model. Again we see from the arguments of the
previous
section that the relative entropy is a monotonic function along this line.
Similarly we can consider the line $t=0$ and $l$ ranging through different
values. Again for positive $l$ the relative entropy is a monotonic function
of this variable. The critical line is a curve in this plane,
since the critical temperature $T_c$ should depend on $l$ and one
needs to change $t$ as a function of $l$ to track it.

It is interesting to consider the reduction of  the two-dimensional phase
diagram
associated with the neighbourhood of the tricritical point to
the one-dimensional phase diagram of the critical point.
This latter fixed point is associated with $l=\infty$ and the
crossover from it to the infinite mass Gaussian fixed point at $t=\infty$
lies completely at infinity in the tricritical phase diagram.
In the previous setting the crossover started from a  finite location
because we did not include the tricritical point.
The  reduction  can be achieved as a projection from the tricritical phase
diagram as follows:
For any value of $(t,l)$ we can let both go
to infinity while keeping their ratio constant. The value of $t/l$
parameterizes
points on the line at infinity. Moreover, that projection is realized by
letting
$z$ run to infinity, thus ensuring that the relative entropy increases in the
process.

One can further appreciate the structure of the phase diagram commented on
above in terms of the shape of RG trajectories,
identified with scaling the non-linear scaling field, where the phase
diagram is presented in these coordinates. In the present case,
the family of scaling curves is
$t=c\,l^{\vf}$ for various $c$, with only one parameter given by the
ratio of scaling dimensions of the relevant fields
$\vf = {\Delta_t \over\Delta_l} >1$, called the crossover exponent.
These curves have the property that
they are all tangent to the $t$ axis at the origin and any straight line
$t=a\,l$ intersects them at some finite point,
$l_i={({a\over c})}^{1\over \vf-1}$
and $t_i=a l_i$.
For any given $c$ the values of $l_i$ and $t_i$ increase as $a$
decreases and go to infinity as $a \rightarrow 0$. This clearly shows
that the stable fixed point of the flow is on the line at infinity and, in
particular, its projective coordinate is $a=0$.
The point $a=\infty$ on the line at infinity is also fixed but unstable.
In general, as the overall factor $z$ is taken to infinity
we shall hit some point on the
saparatrix connecting these two points at infinity.

The tricritical flow diagram that includes the separatrix can be
obtained by a projective transformation (see subsection 3.5). It
is essentially of the same form
as that considered by Nicoll, Chang and Stanley [\cite{NCS}],
with the axes such that the tricritical
point is at the origin (figure 3).
\par\vskip1\baselineskip
\input epsf
\epsfxsize=.7\hsize
\centerline{\bf Fig. 3: Tricritical flow diagram showing the
tricritical, critical and}
\centerline{\bf Gaussian FP (with the mean-field crossover exponent
$\varphi = 2$)} 
\par\vskip1\baselineskip
\centerline{\epsffile{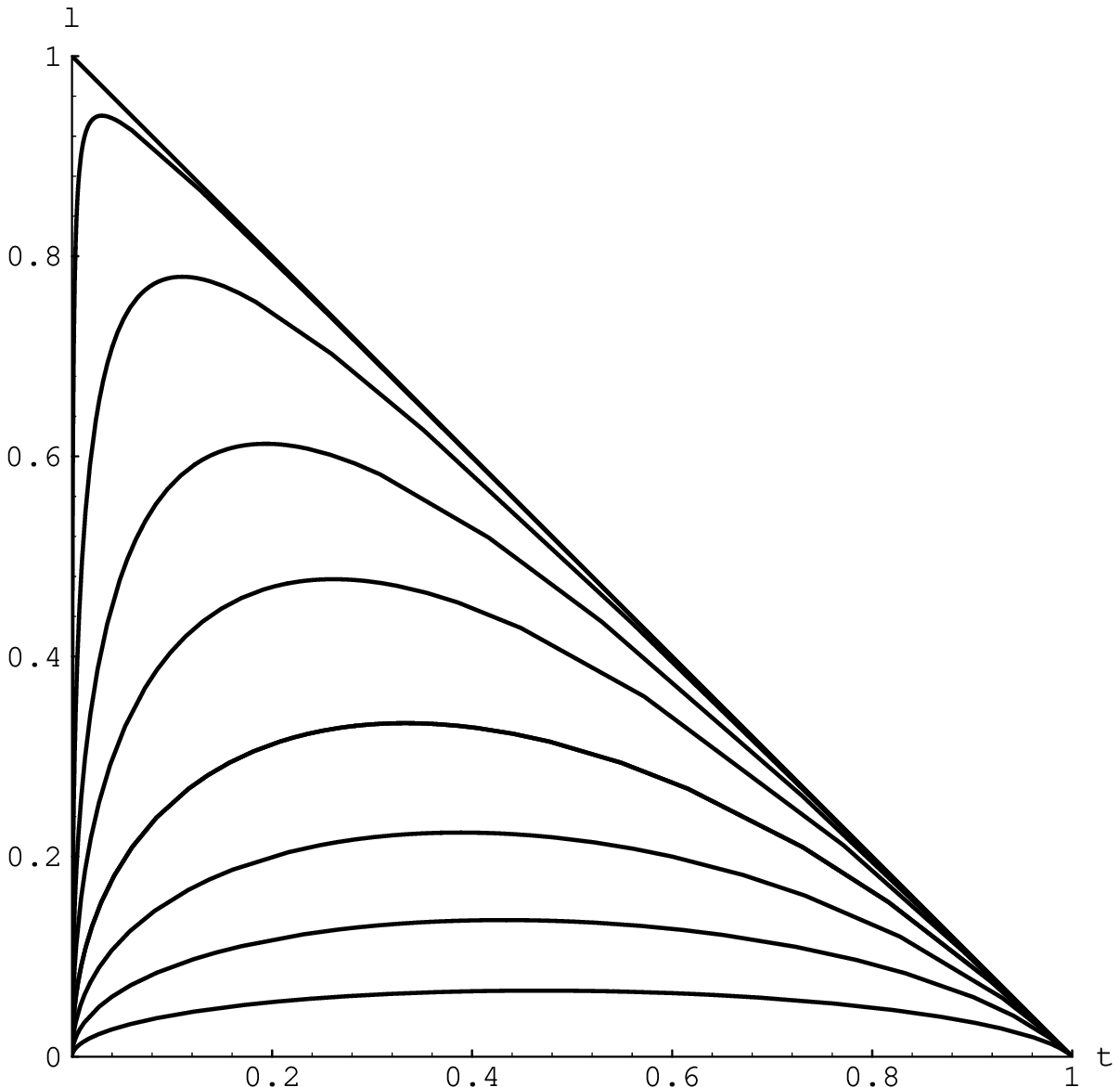}}
\par\vskip1\baselineskip
The critical line is the vertical line (the $l$ axis),
and the crossover to the Gaussian fixed point which is the most stable
fixed point, is the line at infinity, in the positive quadrant of the
$(t,l)$ plane. The Gaussian fixed point is at the end of the horizontal $t$
axis.
Our variable $z$ will parameterize radial lines in this $(t,l)$ plane.
As far as the parameter $\alpha$ is concerned, one could
introduce another axis in the phase diagram,
corresponding to this variable. This can be done for every
crossover, and corresponds to crossover as the momentum is varied.

\subsec{The general case of many crossovers}
The question arises as to the {\it naturalness} of the choice
of a priori distribution $\P_0$. In the case of $Z_2$ models in dimension
$4>d>2$
there is a natural choice
for $\P_0$. It is that field theory with the maximum polynomial potential
that is super-renormalizable in this dimension.
This theory admits the maximum number of non-trivial  universal crossovers
in  this dimension.
For this range of dimensions we therefore choose
$$I_k^0[\phi,\{\l\}]= \int_{\M}\left\{{\alpha\over2}(\partial\phi)^2
+\sum_{a=1}^{k+1}{\l_{2a}\over (2a)!}\phi^{a}\right\}\no$$
and the  full action is then
\eqlabel{\action}
$$I_k[\phi,\{\l\},l_2,\dots,l_{2k}]=I_k^0[\phi,\{\l\}]
+\int_{\M}\left\{\sum_{n=1}^{k}{l_{2n}\over (2n)!}\phi^{2n}\right\}\no$$
The different crossover lines from the multicritical point
can then be arranged to correspond to
flows from the origin along straight lines
(in particular, the coordinate axes).
{}From the general arguments of the previous section the
relative entropy increases along those trajectories.

The crossovers in the above system can be organized in a natural
hierarchical sequence, descending from any one multicritical fixed point
to the one just below in order of criticality.
In this way one loses one irrelevant coupling at each step.
The reduced phase diagram at each step is the hyper-plane at infinity of the
previous diagram. Thus with, our compactification, they constitute a sequence
of
nested projective spaces, ending in a point. This structure deserves
a more detailed treatment.

\subsec{The Geometrical Structure of the Phase Diagram}
The phase diagrams for the critical models corresponding to different
RG fixed points are nested in a natural way as projective spaces,
$${\rm R}P_k \supset {\rm R}P_{k-1}  \supset \cdots \supset {\rm R}P_1
\supset {\rm R}P_0,$$ with ${\rm R}P_0$ being just a point that
represents the infinite mass Gaussian fixed point.
In the action (\docref{action}) the set of couplings $l_{2n}$ together with
the coupling $\lambda_{2k+2}$ lend themselves to an interpretation
as homogeneous coordinates for the projective space ${\rm R}P_k$.
 The value of $\lambda_{2k+2}$ is to be held fixed along any crossover so that
the ratios $r_{2n} = {l_{2n}
\over \lambda_{2k+2}}$ become affine coordinates.
Moreover, in the crossover
from an upper critical point to a lower critical point,
e.g. the tricritical to critical crossover,  the phase diagram for the latter
is realized as the codimension-one (hyper)plane at infinity,
which is equivalent to $\lambda_{2k+2} = 0$. Thus $\lambda_{2k+2}$
effectively disappears from the action of the next critical point,
which has $l_{2k}$ as the highest coupling in the sequence.
The set of couplings $l_{2}, \dots, l_{2k}$ then constitute a
system of homogeneous coordinates in the reduced phase diagram.
One can reach a point of this phase diagram by making $z$ go to infinity
for different (fixed) values of  $l_{2i} \over l_{2k}$. This realization
ensures that
the relative entropy of points in this second phase diagram is
lower than that of points of the first
via monotonicity in $z$ as discussed earlier.

One might, however, think that both phase diagrams cannot be incorporated
in the same picture. This is not so:
One can  perform
a projective change of coordinates so as to bring
the (hyper)plane at infinity to a finite distance. This
can be achieved by first rescaling to $\lambda_{2k+2} = 1$.
For example, in the tri-critical to critical crossover of $\S$3.2,
the condition that $g$ be fixed (e.g. $g=1$ where we now use dimension-less
couplings, the original $g$, which we now label $g_B$, setting the scale)
places
the phase diagram of the critical fixed point at infinity.
However, new homogeneous coordinates $\bar r$ and $\bar\l$ and $\bar g$,
defined so that the projective space is realized as the plane $r + \l + g = 1$
rather than by $g=1$, can be specified by defining
$$\eqalign{{\bar r}&= r\cr
{\bar\l}&= \l \cr
{\bar g}&= r + \l + g.} \no$$
In these co-ordinates our previous ratios, that is, the affine coordinates,
take the form,
$$\eqalign{
{r\over g}&=
{{\bar r\over{\bar g}} \over 1-{\bar r\over {\bar g}}-{\bar \l\over {\bar g}}},
\cr
{\l\over g}&=
{{\bar\l\over{\bar g}} \over 1-{\bar r\over {\bar g}}-{\bar \l\over {\bar g}}}.
}\no$$
The phase diagram in the new co-ordinates, drawn
in figure 3, is patently compact.
Transformations of the this type have been used before in global studies of
the RG [\cite{NCS}].
Another possible realization of the phase diagram would be
to project onto the plane $\l + g = 1$. The new coordinates are
given by
\eqlabel{\compactcoords}
$$\eqalign{
{r\over g}&=
{{\bar r\over{\bar g}} \over 1-{\bar \l\over {\bar g}}},
\cr
{\l\over g}&=
{{\bar\l\over{\bar g}} \over 1-{\bar \l\over {\bar g}}}.
}\no$$
The resulting projective coordinate change converts the line at infinity
into the line $\l = 1$. The  critical fixed point  is on this line at $r=0$ but
the infinite mass Gaussian point remains at $r=\infty$.
Hence we can identify the resulting phase diagram as that of the critical
model.
Similar considerations apply quite generally to the entire hierarchy.

We see that the new ratios in (\docref{compactcoords}) resemble
the solution of typical one loop RG equations.
This is not necessarily accidental.
In practice when one goes from bare to renormalized co-ordinates one
defines the new co-ordinates in terms of normalization
conditions \citelabel{\EnvfRG}[\cite{EnvfRG}], which
can be chosen so that the range of these renormalized co-ordinates
ranges over a finite domain, e.g. from zero to the fixed point value of
the renormalized coupling.  For example, in the $\phi^4$ model
the relation between bare and renormalized couplings at one loop
is given by $$\lambda_{b}={\lambda_{r}\over
1-a(d)\,\lambda_{r}\,R^{4-d}}$$
with $R$ the IR cutoff and
$a(d)$ a dimension dependent factor. If terms of the dimension-less
couplings $a(d)\,\lambda\,R^{4-d}$ we have precisely (\docref{compactcoords}).
However, at higher order in the loop expansion
such normalization conditions may
realize the projective space of the phase diagram in a more
complicated fashion than (\docref{compactcoords}). Nevertheless, one can
think of the change from ``bare'' to renormalized co-ordinates as
the transition from affine co-ordinates to a realization of the projective
space.

\beginsection{Wilson's RG and entropy growth}
Field theoretic renormalization groups that are based on
reparameterization of the couplings 
are a powerful tool for the study of crossovers and the
calculation of
crossover scaling functions, as discussed in [\cite{EnvfRG}]. In essence they can be viewed as implementing
appropriate projective changes of co-ordinates implied by the above discussion.
We now wish to discuss the relative entropy in a Wilsonian context.
A Wilson RG transformation is such
that it eliminates degrees of freedom of short wave length and hence
high energy. Typical examples are decimation or block spin transformations.
It is intuitively clear that their action discards information on
the system and therefore must produce an increase of entropy.
Indeed, as remarked by Ma \citelabel{\Ma}[\cite{Ma}] iterating this type
transformation does not constitute a group but rather a semi-group, since
the process cannot be uniquely reversed. In the language of statistical
mechanics we can think of it as an irreversible process.

For concreteness we illustrate our approach by a very simple example,
the  Gaussian model with action
$$I={1\over2}\int_{0}^{\Lambda}d^dp\,\,\phi(p)\,\left(p^2+r\right)\,\phi(-p),\no$$
which yields
$$W[z]={1\over2}\int_{0}^{\Lambda}{d^dp\over
{(2\pi)}^d}\,\ln{p^2+r\over\Lambda^2}.
\no$$
This model has been already considered in subsection 3.1 but with a
lattice cutoff instead of a momentum cutoff. The relevant coupling that
effects the crossover is $z = t = r-r_c$. The corresponding relative 
entropy $$S[z]={1\over2}\int_{0}^{\Lambda}{d^dp\over(2\pi)^d}
\left( \ln{p^2+r\over {p^2+r_c}} - {t\over p^2+r}\right)\no$$ is finite when 
$\Lambda$ 
goes to infinity, agreeing with (\docref{gaussrelent}),
 and vanishes for $t=0$. The Wilson RG is 
implemented by letting $\Lambda$ run to lower values. Let us see that $S$
is monotonic with $\Lambda$.

We have that $${\partial S\over\partial\Lambda} = {\Lambda^{d-1}\over
2^d \pi^{d\over 2}\, \Gamma(d/2)} \left(
\ln{\Lambda^2+r\over {\Lambda^2+r_c}} - {t\over \Lambda^2+r}\right),\no$$
except for an irrelevant constant. With the change of variable $x =
\Lambda^2$, we have to show that the corresponding function of $x$ is
of the same sign everywhere. Then we want $$\ln{x+r\over x+r_c} - {r-r_c\over
x+r}$$ not to change sign. Interestingly, the properties of this
expression are independent of $x$ somehow for if one substitutes in
$\ln\rho - {\rho -1 
\over\rho}$ the value $\rho = {x+r\over x+r_c}$ then one recovers the entire
function. 
Now it is easy to show that $\ln\rho \geq 1 - {1\over\rho}.$ (The
equality holds for $\rho =1$---the critical point.) 
This proof resembles somehow the classical proofs of $H$-theorems.

We plot in figure 4 the associated relative entropy for this model as a
function of $\Lambda$ to show that it is again a monotonic function.
This behavior is actually closely related to the monotonicity with $r$
considered before: The relative entropy as well as $W$ is a function of
the ratio $r\over\Lambda^2$, which is precisely the solution of the RG
for this simple model. 
\par\vskip1.5\baselineskip
\input epsf
\epsfxsize=0.6\hsize
\centerline{\bf Fig. 4: The Wilsonian Relative Entropy of the Gaussian Model}
\par\vskip.5\baselineskip
\centerline{\epsffile{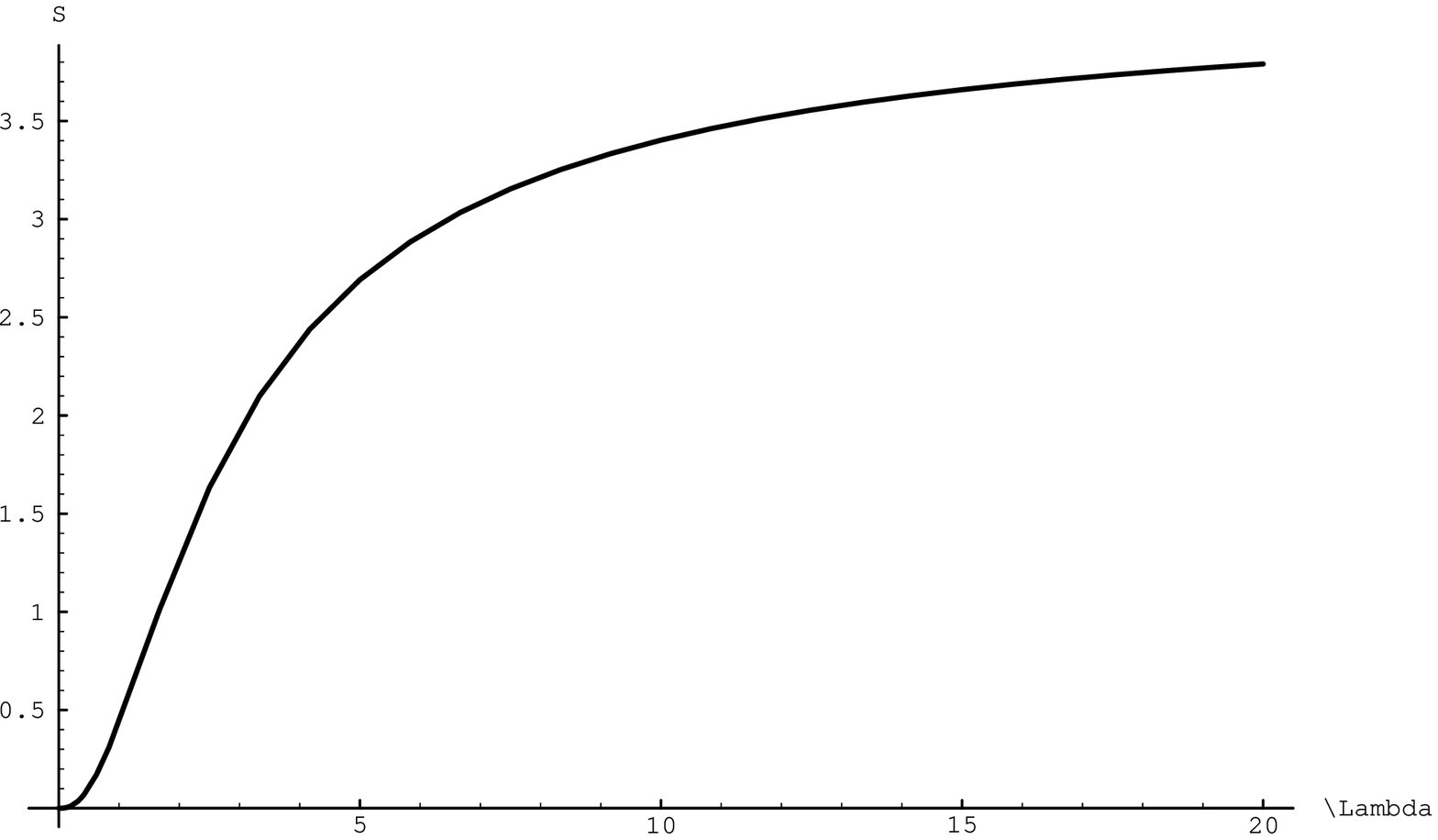}}
\par\vskip1\baselineskip

There are certain features common to all formulations
of Wilsonian RGs for a generic model.
Even if the theory is simple at the scale of the cutoff, as may happen
when we use a lattice model as our regularized theory, a Wilson RG
transformation
complicates it by introducing new couplings. Thus the action of
Wilson's RG is defined in what is called theory space,
typically of infinite dimension,
comprising all possible theories generated by its action.
In practice, one is interested in the critical behavior controlled by
a given fixed point and the theory space reduces to
the corresponding space spanned by the marginal and relevant operators.
Under the action of the RG, the irrelevant coupling constants approach values
which are functions of the relevant coupling constants. In the language
of differential geometry, the RG flow converges to a manifold parameterized
by the relevant couplings, called the critical manifold.
Therefore, the information about the original trajectory
or the value of the couplings at the scale of the cutoff is lost.
In the language of FT, we can say that the non-renormalizable
couplings vanish (or, in general, approach predetermined values)
when the cutoff is removed \citelabel{\Polchinski}[\cite{Polchinski}].

As described above, the action of the Wilson RG is reminiscent of the course of
a typical non-equilibrium process in statistical physics. The initial
state may be set up to be simple but if it is not in equilibrium
then it evolves, getting increasingly complicated until
an equilibrium state is reached, where the system can be described
by a small number of thermodynamic variables. This idea can be
rigorously formulated as Boltzmann's $H$ theorem. In the modern version
of this theorem \citelabel{\Jaynes}[\cite{Jaynes}] $H$ is a function(al)
of the probability distribution of the system defined as
$H=-\cS_a$ of (\docref{Entropy}). It measures the information available to
the system and has to be a minimum at equilibrium. To be precise, the
actual probability distribution
is such that it does not contain information other than that implied
by the constraints or boundary conditions imposed at the outset.

The simplest case of the $H$ theorem is when there is no constraint
wherein $H$ is a minimum for a uniform distribution. This is sometimes called
the principle of equiprobability.
{From a philosophical standpoint, it is based in
the more general principle of sufficient reason, introduced by Leibnitz.
In our context, it can be quoted as stating that if to our knowledge
no difference can be ascribed to two possible outcomes of an aleatory
process, they must be regarded as equally probable.}
This is the case for an isolated system in statistical mechanics;
all the states of a given energy have the same probability
(micro-canonical distribution). Another illustrative example,
is provided by a system  thermally coupled
to a heat reservoir at a given temperature where we want to impose
that the average energy takes a particular value.
Minimizing $H$ then yields the canonical distribution.

In general, we may impose constraints on a system with states $X_i$ that
the average values of a set of functions of its state, $f_r(X_i)$,
adopt pre-determined values,
$$
\langle f_r\rangle := \sum_i P_i\, f_r(X_i) = \bar f_r,
$$
with $P_i := P(X_i)$. The maximum entropy formalism leads to
the probability distribution \citelabel{\Tribus}[\cite{Tribus}]
$$
P_i =  Z^{-1}\,\exp {\left(-\sum_r \lambda_r\,f_r(X_i) \right)}.
$$
The $\lambda_r$ are Lagrange multipliers determined in terms
of $\bar f_r$ through the constraints.
In field theory a state is defined as a field configuration $\phi(x)$.
One can define functionals of the field ${\cal F}_r[\phi(x)]$.
These functionals are usually quasi-local and are called composite
fields. The physical input of a theory can be given in two ways, either by
specifying the microscopic couplings or by specifying the expectation
values of some composite fields, $\langle {\cal F}_r[\phi(x)]\rangle$.
The maximum entropy condition provides an expression
for the probability distribution,
$$
P[\phi(x)] = Z^{-1}\,
{\rm exp} {\left(-\sum_r \lambda_r\,{\cal F}_r[\phi(x)] \right)},
$$
and therefore for the action,
$$
I = \sum_r \lambda_r\,{\cal F}_r;
$$
namely, a linear combination of relevant fields with coupling constants
to be determined from the specified $\langle {\cal F}_r\rangle$.

The formulation of the $H$ theorem described above is very general.
The situation that concerns us here is the
crossover from the critical behavior in the vicinity of
a multicritical point to another more stable multicritical point
under the action of the RG. As soon as a relevant field takes
a non-vanishing value, the action of the RG drives the system
away from the first fixed point towards the second.
In our hierarchical sequence of critical points this was achieved by
the couplings being sent to infinity relative to one another
in a fashion that  descended along this hierarchy.
As described above, the condition represented by fixing the expectation
value of 
the relevant field can be understood as imposing a constraint via
the introduction of a Lagrange multiplier which appears as a coupling
$\lambda_i$
in the field theory.  As in the case of  the introduction
of $\beta$ (inverse temperature), when $\lambda_i$ is sent to infinity we
expect the
entropy to decrease and thus our relative entropy should increase.
Conversely, releasing the constraint is equivalent to  sending the
coupling to zero and the relative entropy decreases.
In the above description the underlying theory is held fixed and only
one parameter varied as one moves through a sequence of ``quasi-static''
states. 

In the Wilson RG picture certain
expectation values are held fixed while the microscopic theory is
allowed to evolve. 
This involves the crossover from cutoff dependent degrees of freedom
to cutoff independent ones and generically falls into the non-equilibrium
situation described above. In this process
one expects that the entropy will actually increase as the system evolves.
This means that our relative entropy should decrease. One can easily see
from figure 4 in the example described at the beginning of this section that
this is indeed the case. In terms of renormalized couplings for given
values of the couplings, we can start with any
value of $\l_i$ and let the RG act. All the trajectories converge
to the critical manifold where $\l_i$ is determined by
the other couplings, $\l_i\left(\l_r\right)$.
The trajectories approach each other in a sort of reverse chaotic
process. In a chaotic process there is great sensitivity to the initial
conditions,
however, in the RG flow there is great insensitivity to the initial
values of the irrelevant couplings which diminish as the flow progresses
and in fact vanish at the end of the flow.

\beginsection{Conclusions}
We have established that the field theoretic relative entropy provides a
natural
function which ranks the different critical points in a model. It grows
as one descends the hierarchy in the crossovers between scalar field
theories corresponding to different multicritical points.
This is a consequence of general properties of the entropy and, in
particular, of the relative entropy.

We have further established that the phase diagrams of the hierarchy of
critical points are associated with a nested sequence of projective
spaces. It is convenient to use coordinates adapted to a particular
phase diagram in the hierarchy. Hence a crossover implies a coordinate
change. The transition from bare to renormalized co-ordinates provides a
method of compactifying the phase diagram. By changing from the
bare coordinates, in which the phase diagram naturally ranges over
entire hyper-planes to appropriate renormalized ones the phase diagram
can be rendered compact.

We discussed the action of the Wilson RG and argued that
the relative entropy increases as more degrees of freedom are
integrated out, when the underlying Hamiltonian is held fixed.  However, when
the
Hamiltonian is allowed to flow, as it generically is in a Wilson
RG, the resulting flow corresponds to a non-equilibrium process in
thermodynamics. Nevertheless, the general formulation of
the $H$-theorem provided by Jaynes allows us to conclude that
the entropy increases in such a process and that the relative entropy
(due to our choice of signs) decreases. In contrast,
the field theoretic crossover
wherein one moves from one point in a phase diagram to another
by varying one of the underlying parameters (such as temperature)
corresponds to a sequence of quasi-static states and in the case
of our hierarchical sequence as one descends the sequence by
sending various parameters to infinity one is gradually placing
tighter constraints much as reducing the temperature does in the canonical
ensemble.
Thus one expects the entropy should reduce and the relative entropy
increase. This is indeed what we find.

One might wonder as to the connection between our entropy
function and the Zamolodchikov $C$ function. It is unlikely that
in two dimensions the two are the same.
Zamolodchikov's $C$ function is built from correlation data and
in the case of a free field theory it is easy to check that the
two functions do not coincide.

{\bf Acknowledgements:} The authors express their gratitude to the
Institute for Physics, University of Amsterdam, where much of this work
was done, for its hospitality and financial support. Denjoe O'Connor
is grateful to John Lewis for several helpful conversations. J.~Gaite
acknowledges support under grant PB 92-1092.
\vskip 10 true pt

\centerline{\bf REFERENCES}

\item{[\cite{Fisher}]}M.E. Fisher, {\it Phases and Phase Diagrams: Gibbs's
Legacy Today}
in {\it Proceedings of the Gibbs Symposium} Yale University, 1989,
American Mathematical Society  (1990) ed. D.G. Cali and G. D. Mostow.
\item{[\cite{various}]}D.J. Wallace and R.K.P. Zia, Ann. of Phys. {\bf
92} (1975) 142; J.L. Cardy, Phys. Lett. {\bf B 215} (1988)
749; A. Cappelli, D. Friedan and
J. Latorre, Nucl. Phys. {\bf B 352} (1991) 616;
A. Cappelli, J.I. Latorre and X. Vilas{\'\i}s-Cardona, Nucl. Phys.
{\bf B 376} (1992) 510; V. Periwal, Mod. Phys. Lett. {\bf A 10} (1995) 1543
\item{[\cite{Balian}]}R. Balian, {\it From Micro-physics to
Macro-physics}, Vol 1, Monographs in Physics, Springer-Verlag 1991.
\item{[\cite{Merca}]}J. P{\'e}rez-Mercader, ``Quantum fluctuations and
irreversibility'' in {\it The Physical Origins of Time Asymmetry},
Eds. J.J. Halliwell, J. P{\'e}rez-Mercader and W. Zurek, Cambridge
University Press (1994) 
\item{[\cite{BombSorkSred}]}L.Bombelli, R.Koul, J. Lee and R. Sorkin, {\it
Phys. Rev.}
{\bf D34} (1986) 373. M. Srednicki, {\it Phys. Rev. Lett.} {\bf 71} (1993) 666.
\item{[\cite{Hol}]} C. Holzhey, F. Larsen and F. Wilczek, {\it Nucl. Phys.}
{\bf B424 [FS]} (1994) 443.
\item{[\cite{Ellis}]} Richard S. Ellis, {\it Entropy, Large Deviations and
Statistical Mechanics}, Springer Berlin (1995)
\item{[\cite{LewisPfisterSullivan}]}J.T. Lewis, C.E. Pfister and W.G. Sullivan,
J. Stat. Phys. {\bf 77} (1994) 397.
\item{[\cite{LewisPfister}]}J.T. Lewis and C.E. Pfister, {\it Thermodynamic
Probability Theory:
Some Aspects of Large Deviations}, DIAS-STP-93-33, to appear in Russian
Mathematical Surveys (Uspekhi) dedicated to A.N. Kolmogorov.
\item{[\cite{Amari}]}Shun-ichi Amari, {\it Differential-Geometric Methods in
Statistics},
Lecture Notes in Statistics {\bf 28}, Springer-Verlag (1985).
\item{[\cite{OConSte}]}
Denjoe O'Connor and C.R. Stephens, {\it Geometry, The Renormalization Group and
Gravity},
in {\it Directions in General Relativity}, Proceedings of the 1993
International Symposium,
Maryland, Vol 1, Cambridge University
Press, (1993),
edited by B. L. Hu,  M.P. Ryan and C. V. Vishevshawara.
\item{[\cite{Wegner}]}F.J. Wegner, Phys. Rev. {\bf B 5} (1972) 4529
\item{[\cite{Arnold}]}V.I. Arnold, {\it Chapitres suppl{\'e}mentaires de
la th{\'e}orie des {\'e}quations diff{\'e}rentielles ordinaires}, MIR, Moscow
(1980)
\item{[\cite{Lefschetz}]}S. Lefschetz, {\it Differential Equations:
Geometric Theory}, Pure and Applied Mathematics--Vol V1, Interscience
Publishers (1962). 
\item{[\cite{BerlinKac}]}T.H. Berlin and M. Kac, {\it Phys. Rev.} {\bf 86}
(1952) 821.
\item{[\cite{Onsager}]}L. Onsager, Phys. Rev. {\bf 65} (1944) 117.
\item{[\cite{NCS}]} J.F. Nicoll, T.S. Chang and H.E. Stanley,
Phys. Rev. {\bf B 12} (1975) 458
\item{[\cite{EnvfRG}]}Denjoe O'Connor and C.R. Stephens,
IJMP, {\bf A9}, (1994), pp. 2805-2902.
\item{[\cite{Ma}]}S. Ma, {\it Modern Theory of Critical Phenomena},
Frontiers in Physics 1982.
\item{[\cite{Polchinski}]}J. Polchinski, Nucl. Phys. {\bf B 231} (1984) 269
\item{[\cite{Jaynes}]}E.T. Jaynes, {\it Papers on Probability Theory,
Statistics and Statistical Physics}, D. Reidel Pub. Co., Dordrecht (1983)
\item{[\cite{Tribus}]}M. Tribus, {\it Thermostatics and Thermodynamics},
D. Van Nostrand Co., Princeton (1961)
\bye